# Observation of quantum oscillations near the Mott-Ioffe-Regel limit in CaAs$_3$

Yuxiang Wang[1,#], Minhao Zhao[2,#], Jinglei Zhang[4,#], Wenbin Wu[5], Shichao Li[6], Yong Zhang[4], Wenxiang Jiang[7], Nesta Benno Joseph[8], Liangcai Xu[9], Yicheng Mou[1], Yunkun Yang[2], Pengliang Leng[2], Yong Zhang[10], Li Pi[4], Alexey Suslov[11], Mykhaylo Ozerov[11], Jan Wyzula[12], Milan Orlita[12], Fengfeng Zhu[7], Yi Zhang[13], Xufeng Kou[10], Zengwei Zhu[9], Awadhesh Narayan[8], Dong Qian[7], Jinsheng Wen[6], Xiang Yuan[5,14,*], Faxian Xiu[1,2,3,15,16,*], Cheng Zhang[1,15,*]

[1] State Key Laboratory of Surface Physics and Institute for Nanoelectronic Devices and Quantum Computing, Fudan University, Shanghai 200433, China

[2] State Key Laboratory of Surface Physics and Department of Physics, Fudan University, Shanghai 200433, China

[3] Shanghai Qi Zhi Institute, 41st Floor, AI Tower, No. 701 Yunjin Road, Xuhui District, Shanghai, 200232, China

[4] Anhui Province Key Laboratory of Condensed Matter Physics at Extreme Conditions, High Magnetic Field Laboratory of the Chinese Academy of Sciences, Hefei 230031, China

[5] State Key Laboratory of Precision Spectroscopy, East China Normal University, Shanghai 200241, China

[6] National Laboratory of Solid State Microstructures and Department of Physics, Nanjing University, Nanjing 210093, China

[7] Key Laboratory of Artificial Structures and Quantum Control (Ministry of Education), School of Physics and Astronomy, Shanghai Jiao Tong University, Shanghai 200240, China

[8] Solid State and Structural Chemistry Unit, Indian Institute of Science, Bangalore 560012, India

[9] Wuhan National High Magnetic Field Center and School of Physics, Huazhong University of Science and Technology, Wuhan 430074, China

[10] School of Information Science and Technology, ShanghaiTech University, Shanghai 201210, China

[11] National High Magnetic Field Laboratory, Tallahassee, Florida 32310, USA

[12] LNCMI-CNRS UPR3228, Université Grenoble Alpes, Université Toulouse 3, INSA Toulouse, EMFL 25 rue des Martyrs, BP166, Grenoble Cedex 9 38042, France

[13] International Center for Quantum Materials, School of Physics, Peking University, Beijing 100871, China

[14] School of Physics and Electronic Science, East China Normal University, Shanghai 200241, China

[15] Zhangjiang Fudan International Innovation Center, Fudan University, Shanghai 201210, China

[16] Shanghai Research Center for Quantum Sciences, Shanghai 201315, China

[#] These authors contributed equally to this work

[*] Correspondence and requests for materials should be addressed to X. Y. (E-mail: xyuan@lps.ecnu.edu.cn), F. X. (E-mail: Faxian@fudan.edu.cn) & C. Z. (E-mail: zhangcheng@fudan.edu.cn)

**Abstract**

The Mott-Ioffe-Regel limit sets the lower bound of carrier mean free path for coherent quasiparticle transport.[1,2] Metallicity beyond this limit is of great interest because it is often closely related to quantum criticality and unconventional superconductivity.[3–5] Progress along this direction mainly focuses on the strange-metal behaviors originating from the evolution of quasiparticle scattering rate such as linear-in-temperature resistivity[4], while the quasiparticle coherence phenomena in this regime are much less explored due to the short mean free path at the diffusive bound. Here we report the observation of quantum oscillations from Landau quantization near the Mott-Ioffe-Regel limit in $CaAs_3$. Despite the insulator-like temperature dependence of resistivity, $CaAs_3$ presents giant magnetoresistance and prominent Shubnikov–de Haas oscillations from Fermi surfaces, indicating highly-coherent band transport. In contrast, the quantum oscillation is absent in the magnetic torque. The quasiparticle effective mass increases systematically with magnetic fields, manifesting a much larger value than the expectation given by magneto-infrared spectroscopy. It suggests a strong many-body renormalization effect near Fermi surface. We find that these unconventional behaviors may be explained by the interplay between the mobility edge and the van Hove singularity, which results in the formation of coherent cyclotron orbits emerging at the diffusive bound. Our results call for further study on the electron correlation effect of the van Hove singularity.

**Main text**

As described by the Boltzmann theory, diffusive electron transport lies on the basis of long-lived quasiparticles, which treats carriers as electron gases moving semiclassically between positively charged ionic cores. The interaction effects with Bloch lattice potential as well as other quasiparticles are enclosed in the renormalized effective mass. The long-lived quasiparticle picture is only valid when the mean free path $l$ exceeds the Fermi wavelength. It sets the Mott-Ioffe-Regel (MIR) limit ($k_F l > 1$ with $k_F$ the Fermi wave vector) for metals, below which the quasiparticle is no longer well-defined[1] and a metal-to-insulator transition occurs due to electron localization.[6,7] In recent years, observation of metallicity beyond the MIR limit, especially the strange metal behavior in strongly correlated materials, challenges such a transitional point of view.[3,5,8–13] Owing to its intimate connection with quantum criticality and unconventional superconductivity, metallicity beyond the MIR limit has been intensively investigated but remains mysterious.

Apart from the temperature dependence of resistivity, the presence of a Fermi surface formed by mobile electrons is another important characteristic of metals. When subjected to an external magnetic field, electrons experience Landau quantization and form cyclotron orbits along the contour of a closed Fermi surface. It leads to a series of magnetic quantum oscillations in conductivity, magnetization and thermoelectricity, which have been established as one of the most powerful tools to characterize the Fermi surface properties.[14] In order to generate Landau quantization, electrons should remain phase coherent during the cyclotron period. Consequently, quantum oscillations are typically observed in clean metallic systems with high carrier mobility and long mean free path, including but not limited to two-dimensional electron gases[15] and topological materials[16,17]. With the reduction of the mean free path, electrons gradually lose coherence upon scattering within the cycle of an orbit, which eventually eliminates the quantum oscillations at finite temperatures. Beyond the MIR limit, electrons generally become incoherent and are not expected to present quantum oscillations[1,2,18].

Here we report the observation of quantum oscillations and quasiparticle mass enhancement in the vicinity of the MIR limit in CaAs$_3$. CaAs$_3$ belongs to the CaP$_3$ family of materials, which were predicted to be topological nodal line semimetals in the absence of spin-orbit coupling.[19,20] The inclusion of spin-orbit coupling opens a small gap in the nodal loop. No sign of topological surface states was found in previous photoemission experiments[21]. We performed magneto-transport experiments on CaAs$_3$ bulk crystals in variable temperature inserts with a superconducting magnet (9 T), a water-cooled resistive magnet (38 T), and a hybrid magnet (45.22 T). The crystal surface of the as-grown CaAs$_3$ samples is determined to be the (010) plane using X-ray diffraction as shown in Fig. S1a. The current was applied in the (010) plane with the magnetic field $B$ perpendicular to the plane.

Fig. 1a shows the temperature $T$ dependence of the longitudinal resistivity $\rho_{xx}$ in Sample C1 at zero magnetic field. As $T$ changes from 360 K to 2 K, $\rho_{xx}$ increases exponentially by about three orders of magnitude, showing an insulator-like $T$ dependence similar to previous reports[21]. We employ the Arrhenius equation $\rho = \rho_0 e^{-\Delta/k_B T}$ to quantify the thermal activation behavior of $\rho_{xx}$. Here $\Delta$ and $k_B$ are activation energy and the Boltzmann constant, respectively. As illustrated in Fig. 1b, two regimes with distinct activation energies ($\Delta_1$=147 meV for 220~360 K, $\Delta_2$=5.89 meV for 20~60 K) can be observed in the Arrhenius plot ($\ln \rho_{xx}$ vs. $1/T$). Below 20 K, $\rho_{xx}$ gradually deviates from the activation behavior and becomes saturated towards the low-temperature limit (refer to Fig. S1b for the $\rho_{xx}$-$T$ curve of Sample C2 down to 50 mK). According to the heat capacity data shown in Fig. S1c, no sign of any phase transition is detected in the range of 10~150 K. To help understand the thermal activation of $\rho_{xx}$, we measured the bulk electronic structure and band filling of CaAs$_3$ crystals by angle-resolved photoemission spectroscopy (ARPES). The dispersion along the $\Gamma - X$ direction at 10 K is shown in Fig. 1c. No significant variation in the band dispersion is found as the temperature increases to 45 K and 70 K (Fig. S2). The Fermi level is located at roughly 42 meV above the conduction band edge along with a gap of 158 meV, which is also consistent with previous ARPES results obtained in CaAs$_3$[21]. It suggests that the thermal activation from the valence band to the conduction band accounts for $\Delta_1$ in the range of 220~360 K.

We then present Hall resistivity $\rho_{yx}$ as a function of $B$ in Fig. 1d. $\rho_{yx}$ is linear in $B$ at all temperatures, suggesting one kind of carriers dominates the transport. Surprisingly, a sign reversal of the Hall slope is observed at around 210 K. By performing linear fitting to the $\rho_{yx}$-$B$ curves, we extract the Hall coefficient $R_H$ as shown in Fig. 1e, indicating the carrier changing from $n$- to $p$-type as $T$ decreases. As shown in Fig. S3, $R_H$ also shows similar activation behavior with $\rho_{xx}$, which indicates that the activation of resistivity comes from thermally activated carriers. We further measured the Seebeck coefficient (the inset of Fig. 1e), which is consistent with the Hall effect and shows similar sign reversal at around 220 K. Note that the $p$-type carriers given by the Hall and Seebeck effects at low temperatures contradict the position of Fermi level that intersects the conduction band as observed in ARPES, which will be discussed later. Fig. 1f shows the carrier density $n_H$ and mobility $\mu$ calculated from the Hall effect from 2.5 K to 180 K. In this $p$-type regime, $n_H$ decreases monotonously from 3.3×10$^{17}$ cm$^{-3}$ at 180 K down to 4.8×10$^{15}$ cm$^{-3}$ at 2.5 K, while $\mu$ shows a dip at 20 K and then increases to 263 cm$^2$/Vs at 2.5 K. The carrier density of CaAs$_3$ is comparable to intrinsic narrow-gap semiconductors such as InSb[22] and Hg$_{1-x}$Cd$_x$Te[23] but its mobility value is much lower. The increase of mobility at low temperatures in Fig. 1f may result from the suppression of phonon scattering as well as electron-electron scattering due to the reduced carrier density.

Despite the relatively low mobility, prominent Shubnikov–de Haas (SdH) oscillations and a giant magnetoresistivity (MR) are observed at high magnetic fields up to 38 T. As shown in Fig. 2a, $\rho_{xx}$ starts to oscillate around 7 T and increases by over two orders of magnitude in the range of 20~33 T at 0.3 K. Fig. 2b and c display the temperature dependence of the $\rho_{xx}$-$B$ curves in sample C1 and C3. The insets of Fig. 2b and c are the extracted SdH oscillations at different temperatures. SdH oscillations decay as temperature rises and gradually smear out above 4 K. We determine the oscillation frequency $F = 23.4$ T and Landau level index $n$ using the Landau fan diagram in Fig. S4b. According to the Onsager relation $F = \frac{\hbar}{2\pi e} A$, the oscillation frequency $F$ is proportional to the cross-sectional area $A$ of the Fermi surface along the direction of $B$. Here $\hbar$ is the reduced Planck constant, and $e$ is the elementary charge. When rotating the magnetic field away from the $z$ direction, SdH oscillations persist as shown in Fig. S6 show characteristics of a three-dimensional (3D) Fermi surface. We resolve the Fermi surface anisotropy as a function of $\theta$ and $\varphi$ in Fig. 2d by tracking the oscillation frequency in both the $z$-$y$ and $z$-$x$ planes. It corresponds to an ellipsoid-like shape with large out-of-plane anisotropy ($k_z \sim 5k_x$) and weak in-plane anisotropy ($k_x \sim k_y$), as a result of two Fermi spheres merging together (Fig. S15). It generally agrees with the theoretical curve (the yellow dashed line in Fig. 2d) obtained from the Fermi surface anisotropy given by density functional theory (DFT) calculations. We calculate the in-plane Fermi wave vector as $k_F = 0.0267$ Å$^{-1}$. Importantly, in stark contrast to the large quantum oscillations and giant MR ratio observed in resistivity, the field dependence of the magnetic torque $\tau$ show neither of these features. In Fig. 2e, we show that de Haas–van Alphen (dHvA) oscillation was also absent as the field rotates. It is contradictory to the general belief that dHvA oscillations are more sensitive to Landau levels than SdH oscillations. In Supplementary Note 3, we exclude the crystal inhomogeneity as the possible origin of the absence of dHvA oscillations. These unusual Landau quantization behaviors in CaAs$_3$ will be discussed in detail later.

SdH oscillations can be quantitatively analyzed using the Lifshitz-Kosevich (LK) formula[14], $\frac{\Delta\rho_{xx}}{\rho_{xx}(B=0)} \propto R_T R_D \cos\left[2\pi\left(\frac{F}{B} + \varphi\right)\right]$. Note that $R_T = \frac{2\pi^2(k_B T/\hbar\omega)}{\sinh(2\pi^2 k_B T/\hbar\omega)}$ and $R_D = e^{-2\pi^2(k_B T_D/\hbar\omega)}$ are thermal and Dingle damping factors with the cyclotron frequency $\omega = eB/m^*$ and Dingle temperature $T_D = \hbar/2\pi k_B \tau_q$. Here $m^*$, $\tau_q$ and $\varphi$ are effective mass, quantum lifetime, and the phase offset, respectively. In Fig. 2f and Fig. S4d, we employ the LK formula to fit the temperature and magnetic field dependence of SdH oscillations. The oscillation amplitude is obtained by subtracting the MR background and normalized by $\rho_{BKG}$. The normalization process is to exclude the influence of large MR background following a recent work[24]. The comparison of different normalization methods is shown in Fig.S5, which only weakly impacts the $m^*$ value. After that, the effective mass $m^*$ at different magnetic fields and Dingle temperature $T_D$ are calculated. Interestingly, $m^*$ systematically increases from $0.74m_e$ at 8.73 T to $2.19m_e$ at 31.25 T (Fig. 2g). To exclude possible influence from MR background subtraction, the increase of $m^*$ with magnetic fields is further verified by the second derivative method and the integral method (refer to Supplementary Note 5). Quasiparticle effective mass is generally determined by the band dispersion as well as interactions with other quasiparticles, which are hard to be modulated by external magnetic fields. Therefore, the enhancement of quasiparticle effective mass is highly nontrivial and has only been observed in a handful of limited cases such as high-$T_c$ superconductors near the quantum critical point and nodal-loop semimetals upon magnetic breakdown.[25–27]

By taking the field-dependent value of $m^*$, the fitting to Dingle damping yields a Dingle temperature of $T_D$ = 3.9 K, corresponding to a quantum lifetime of $\tau_q = 3.1 \times 10^{-13}$ s. Then the quantum mean free path is expressed as $l_q = v_F \tau_q = \frac{\hbar k_F \tau_q}{m^*}$, which evolves from 12.9 nm at 8.73 T to 7.1 nm at 18.35 T (Fig. S4). The inset of Fig. 2g shows the ratio of the in-field mobility to its zero-field value, where the in-field value is extracted from the Hall angle $\tan\theta = \mu B$. The ratio dramatically decreases with the increase of $B$, being consistent with the evolution of $l_q$. For comparison, we also extract mean free path from Hall mobility as $l_t = \hbar k_F \mu/e = 4.6$ nm, which is roughly of the same order with $l_q$, respectively. In Fig. S9, we present the transport results of other CaAs$_3$ samples (C4, C5, and C6), which show consistent transport behaviors of Hall coefficient anomaly and quantum oscillations.

To compare the Landau quantization of CaAs$_3$ in a larger energy range which is less influenced by the many-body interaction, we carried out a magneto-infrared spectroscopy study[28,29]. The relative reflectivity $R_B/R_0$ is measured on the (010) plane in the Faraday geometry (with $B$ perpendicular to the plane), where $R_B$ and $R_0$ denote the reflectivity with and without the applied magnetic field, respectively. The spectra measured in a 17.5 T superconducting magnet are presented in Fig. 3a. A larger field range, up to ~30T, is achieved in a resistive coil, and the collected data is shown in a false-color plot (Fig. 3b). A series of prominent peak-to-dip features (peaks are denoted by deep green dots in Fig. 3a) from interband-Landau-level excitations emerge and evolve with $B$. The observed optical transitions agree with the presence of SdH oscillations in transport and contrast the unusual absence of quantum oscillation in magnetic torque. In our data analysis, we associated the minima of $\frac{d(R_B/R_0)}{d\omega}$[30] with the energies of excitations, as shown in Fig. 3c. All observed interband-Landau-level transitions extrapolate to the same energy in the limit of a vanishing magnetic field. The energy difference between adjacent transitions decreases as a function of the photon energy, thus indicating an unequally spaced Landau levels spectrum. These are typical features of massive Dirac fermions. The corresponding Landau level spectrum reads:

$$E_{\pm n} = \pm\sqrt{2e\hbar|n|Bv_F^2 + (\frac{E_g}{2})^2}, \qquad (1)$$

where $n$ is the integer Landau index. The band structure is characterized by the velocity parameter $v_F$ and energy gap $E_g$. The plus and minus signs denote the Landau levels in the conduction and valence bands, respectively. The selection rule follows $\Delta|n| = \pm 1$ as shown in Fig. 3d, where $T_n$ denotes the $-n \to n+1$ and $-(n+1) \to n$ transitions with the corresponding excitation energy given by:

$$\omega_n(B) = \sqrt{2e\hbar(|n|+1)Bv_F^2 + (\frac{E_g}{2})^2} + \sqrt{2e\hbar|n|Bv_F^2 + (\frac{E_g}{2})^2}. \qquad (2)$$

The white dashed curves shown in Fig. 3c are the best fitting results according to Eq. (2) with all transition indices assigned. Due to the absence of splitting for each interband-Landau-level transition, the band structure is expected to be particle-hole symmetric. The overall Fermi velocity is extracted as $v_F = 2.69 \times 10^5$ m/s. The fitted band gap reaches $E_g \sim 165.5$ meV, which is consistent with the gap value extracted from the ARPES as well as the activation energy $\Delta_1$ from the temperature-dependent resistivity. The energy dispersion can be estimated by $E_\pm(k) = E_0 \pm \sqrt{(\frac{E_g}{2})^2 + (\hbar v_F k)^2}$ where $E_0$ defines the zero energy of the low-energy model. The reproduced band dispersion given by magneto-infrared spectroscopy agrees well with the ARPES result (comparison shown in Fig. S7). It further proves that the Landau quantization comes from the gapped bulk band. As shown in Fig. 3e, the Fermi velocity for each set of optical transitions can be separately fitted. The high-index transitions, e.g. $v_F^{T_2}, v_F^{T_3}$, are close to the single-particle picture (orange lines in Fig. 3e). In contrast, for the low-index transitions, the extracted $v_F^{T_0}, v_F^{T_1}$ as well as $m^*$ deviate away from orange lines. It corresponds to a Fermi velocity renormalization from many-body interaction[31–33] for low-energy excitations. Different from the transport measurement where the many-body effect profoundly renormalizes the effective mass near the Fermi energy, the extracted $m^*$ from magneto-infrared spectroscopy by $m^* = \frac{E_g}{2v_F^2}$ (especially for the high index transition) is less influenced and presents much smaller value which is nearly independent of $B$ (Fig. 3f). This is because $m^*$ from magneto-infrared spectroscopy is mainly determined by the Fermi velocity and energy gap of the band. It corresponds to the electronic state property of the entire band due to the much larger detecting energy range compared to transport. Thus, the extracted mass in the infrared spectrum (also known as Dirac mass in the model) corresponds to the single particle mass of the system (refer to Supplementary Note 1. iii). In contrast, the quasiparticle mass from quantum oscillations is sensitive to band structure and interaction effect near Fermi energy. Therefore, the discrepancy between $m^*$ values from magneto-infrared and magneto-transport experiments suggests the Fermi surface electrons are influenced by a strong renormalization effect, which is also likely to be responsible for the observed mass enhancement in magnetic fields. (Refer to Supplementary Note 1 for more discussion)

Having obtained the quasiparticle properties of CaAs$_3$, we in turn compare it in a scaling plot of Fig. 4 with other bulk systems with quantum oscillations from electronic Fermi surfaces. The product of the Fermi wave vector and mean free path, $k_F l$, has been widely used to determine the metallicity of a material. The MIR limit of $k_F l = 1$ defines the mobility edge, below which the electrons become localized.[1] According to the Drude model, the conductivity is expressed as $\sigma = n\mu e = \frac{e^2}{3\pi^2 \hbar} k_F^2 l$, which results in the ratio of $\sigma/k_F$ scaling linearly with the metallicity parameter $k_F l$. Here the ratio of $\sigma/k_F$ is plotted logarithmically against $k_F l$ over a wide range of 3D bulk materials from high-mobility topological semimetals such as Cd$_3$As$_2$ to narrow-gap semiconductors such as InSb.[34–55] Here the mean free path $l$ is extracted from the Hall mobility. The red dashed line is the theoretical curve given by the Drude model. Note that we mainly focus on materials with simple Fermi surface geometries to avoid complications in $k_F l$ calculation, and materials showing quantum oscillations without bulk electronic Fermi surfaces such as SmB$_6$ and YbB$_{12}$ are also not included[56,57]. Despite the large variation in the carrier density and mobility, the values of $\sigma/k_F$ and $k_F l$ in these materials converge closely around the red dashed line in Fig. 4, and most materials are located in the deep metallic side with $k_F l \gg 1$. In contrast, transport behavior of CaAs$_3$ significantly deviates from the Drude model as temperature decreases and shows a crossover of the MIR limit. At low temperatures where quantum oscillations emerge, the conductivity of CaAs$_3$ is over two orders of magnitude smaller than the value given by the Drude model. It is also worth mentioning that CaAs$_3$ has the smallest $k_F l$ value (~1.2 at 2.5 K) among these materials at the temperatures where quantum oscillations persist.

Now we summarize the major observations in the transport results of CaAs$_3$: (1) the Hall and Seebeck effects indicate *p*-type carriers, while the Fermi level crosses the conduction band edge; (2) strong 3D SdH oscillations with large resistivity appear near the MIR limit, while no dHvA oscillations are observed; (3) the quasiparticle effective mass given by quantum oscillations increases systematically with magnetic fields and is much larger than that extracted from magneto-infrared spectroscopy. These experimental behaviors violate the standard Fermi liquid picture in the Boltzmann theory and are distinct from conventional narrow-gap semiconductors or band insulators. On the other hand, the coherent transport with Landau quantization and SdH oscillations also differs from typical non-Fermi-liquid systems, in which electrons often become incoherent. The metallicity parameter $k_F l$ close to the MIR limit suggests that transport in CaAs$_3$ is near the diffusive bound.

To explain the aforementioned behaviors, we propose a phenomenological model as shown in Fig. 5. Figure 5a is a sketch of the bands and density of states (DOS) in CaAs$_3$. The yellow and light blue regions separated by the mobility edge represent the extended and localized states, respectively, and the orange region corresponds to conducting carriers of the extended states near the Fermi energy $E_F$. Since the metallicity parameter barely crosses the MIR limit, only electrons in a small range around $E_F$ are mobile. In contrast, electrons at lower energies of the conduction band are located in the mobility gap and therefore become localized at low temperatures. The carrier localization is supported by the comparison between the Fermi surface size and the Hall density. The carrier density filling the Fermi sphere can be estimated as $n_{FS} = \frac{k_F^x \cdot k_F^y \cdot k_F^z}{3\pi^2} = 3.8 \times 10^{18}$ cm$^{-3}$, almost three orders of magnitude larger than the Hall density at 2.5 K (4.8×10$^{15}$ cm$^{-3}$). Note that here $n_{FS}$ only represents the volume of Fermi sphere rather than the number of mobile electrons participating in SdH oscillations. The dramatic discrepancy between $n_H$ and $n_{FS}$ suggests that the majority of electrons in the Fermi sphere is not counted in Hall effect as shown in Fig. 5b, which account for the deviation from the Drude model of CaAs$_3$ in Fig. 4. Only electrons beyond the mobility edge (orange region in Fig. 5b) contributes to $n_H$ while the rest of them (blue region) are localized at 2.5 K. It could also fit with the absence of quantum oscillation in torque magnetometry. In contrast to electric transport which directly probes the mobile electrons, the magnetization measurement contains the incoherent contribution from localized electrons, which overwhelms the dHvA oscillations (Refer to Supplementary Note 6 for further discussion and simulation). Meanwhile, Landau quantization in magneto-infrared spectroscopy appears at a much larger energy scale, hence not affected by the mobility edge. The carrier localization may be responsible for the activation regime characterized by $\Delta_2$, which corresponds to the thermal excitation of electrons from the localized state below the mobility edge to the extended state above the mobility edge[58].

Due to the opposite dispersion along Y-Γ and Y-X[20], CaAs$_3$ presents a van Hove singularity at the saddle point Y. Similar to its counterpart in two-dimensional (2D) systems, the van Hove singularity in 3D leads to a local maximum in the DOS but with a finite value rather than a logarithmically diverging one as in 2D.[59] For a narrow energy window above the saddle point, the DOS decreases with the increase of energy. We verify the DOS peak near the Lifshitz transition at the saddle point in $CaAs_3$ by the density functional theory calculation (refer to Methods and Fig. S8). As illustrated in Fig. 5a, the Fermi level of $CaAs_3$ lies slightly above the DOS peak as well as the mobility edge. Due to the low Fermi energy, a small amount of disorder can introduce the mobility edge and lead to a metal-to-insulator transition[58]. Since the Fermi surface electrons are located in a region where the DOS decreases with $E$, the carriers show hole-like transport in thermoelectric measurement, resulting in the positive Seebeck coefficient at low temperatures.[60] As $T$ increases, $E_F$ may go beyond this narrow energy window due to the thermally excited electrons from the valence band and the Seebeck coefficient switches back to negative. Similarly, the sign of the Hall effect is determined by the sign of Fermi velocity[60], which typically changes sign together with the slope of local DOS at the saddle point[61,62] as has been recently observed in various moiré superlattices[63,64]. However, due to the momentum and energy resolution, we do not directly resolve the Lifshitz transition by ARPES in our experiment. Other approaches such as scanning tunneling microscopy and laser-based ARPES may be utilized to study the energy dependence of the Fermi contour in this system.

Fig. 5c depicts the transport dynamics of the extended ($k_F l > 1$) and localized ($k_F l < 1$) states, respectively. In the extended states, electrons move freely in the lattice with a mean free path larger than its de Broglie wavelength. It enables a quantum mechanical treatment of the electron wave function during the transport process, giving rise to various electron-coherence phenomena such as Landau quantization and the Aharonov-Bohm effect. Electrons in the localized states are spatially confined due to the small mean free path. At low temperatures, they do not contribute to transport properties such as resistivity, Hall effect, Seebeck effect and SdH oscillations. Upon thermal activation, these localized electrons may become mobile by hopping among different sites. Unlike band transport, hopping transport does not show characteristics of the Fermi surface. In Fig. S11, we show that neither Mott nor Efros-Shklovskii version of the variable range hopping formula can fully describe the resistivity in the low-$T$ region. The mobility edge picture also explains the large variation of resistivity among different samples despite similar Fermi surface size (refer to Fig. S10), since the resistivity is now mainly determined by the number of mobile carriers in the extended states (orange region in Fig. 5). Despite the absence of metallicity in the temperature dependence of resistivity down to our lowest experimental temperature of 50 mK (Fig. S1b), well-defined quantum oscillations are observed in almost every $CaAs_3$ sample we measured with similar frequencies. As the direct consequence of Landau quantization, the presence of quantum oscillations is the hallmark of mobile electrons from the Fermi surface. These intrinsic mobile electrons at ground states then lead to the saturation of resistivity towards zero temperature in Fig. S11.

Perhaps the most striking finding in our study is the highly-coherent electrons, which are robust enough to form cyclotron orbits, residing in the vicinity of the MIR limit. We note that other materials with small metallicity parameters such as CdTe and HgTe shown in Fig. 4 are mainly narrow-gap semiconductors with low Fermi energy and small $k_F$ value. The resultant small Fermi pocket size reduces the requirement of phase-coherence length to complete the cyclotron orbit. However, distinct from these materials, $CaAs_3$ strongly deviates from the red dashed line in Fig. 4 given by the Drude model with resistivity being nearly two orders of magnitude larger. It means that the Fermi energy may lies close to the mobility gap, leaving the mobile electrons forming coherent cyclotron orbits just above the edge of Anderson localization. This peculiar behavior is reminiscent of Landau levels at finite temperatures in quantum Hall physics. Landau levels of quantum Hall systems are non-dispersive flat bands so that quantum coherence forms close to the delocalization transition. Therefore, we anticipate that the flat band around the saddle point contributes to the observed electron coherence near the mobility edge in $CaAs_3$. Although possible deviation of LK formula may occur near the quantum limit[65,66], the thermal damping factor as we used to extract effective mass is less influenced by the oscillation of Fermi energy. We note that such systematical mass enhancement has not been observed in other systems such as Bi, $ZrTe_5$ and $Cd_3As_2$, when approaching low Landau levels[67–70]. The quasiparticle mass enhancement observed in Fig. 2 may result from the strongly suppressed mobility in magnetic fields, which further shifts the system towards the insulating side of the MIR limit.

While not excluding other possibilities, we find that the phenomenological model proposed here could account for three unconventional observations in $CaAs_3$ mentioned earlier. Further tuning the Fermi energy may be carried out to verify the connection between the carrier sign anomaly and the van Hove singularity. It is worth noting that the proposed scenario of exotic metallic states near the mobility edge is not limited to $CaAs_3$ and is expected to appear in other low-carrier-density systems with the van Hove singularity and mobility edge near the Fermi energy as well.

To conclude, we perform systematic quantum transport and magneto-infrared spectroscopy study of $CaAs_3$ bulk crystals. $CaAs_3$ exhibits an insulator-like temperature dependence of resistivity with a metallicity parameter close to the MIR limit. The saddle point in the band structure results in a sign reversal of Hall and Seebeck coefficients. Robust 3D SdH oscillations are observed in magnetic fields accompanied by quasiparticle mass enhancement, suggesting highly-coherent band transport near the MIR limit. Distinct from the widely studied temperature dependence of resistivity, mobile electrons near the MIR limit observed here manifest the metallic characteristic from the aspect of quasiparticle coherence. These unconventional transport properties go beyond the traditional picture of hopping conduction in the critical regime and call for further study on the electron correlation effect of the van Hove singularity at the diffusive bound.

**Methods**

**Crystal growth and characterization.** Similar to the growth method of $SrAs_3$ in the previous study[52], $CaAs_3$ single crystals were grown by melting a stoichiometric mixture of Ca and As at 850 °C. The melt was held at 850 °C for 10 hours before cooling to 700 °C at the rate of 2 °C per hour. $CaAs_3$ bulk crystals with the *a-c* plane as the cleavage plane were then obtained after naturally cooling down to room temperature. The composition and crystal structure of the as-grown crystals were checked by energy-dispersive X-ray spectroscopy and X-ray diffraction (XRD), respectively. Angle-resolved photoemission spectroscopy (ARPES) experiments were carried out at Beamline 4.0.3 in Advanced Light Source with an energy resolution better than 15 meV. $CaAs_3$ crystals were cleaved *in situ* at 10 K with a base pressure better than $1 \times 10^{-10}$ Torr.

**Transport measurement.** The magneto-transport measurements were performed in commercial variable temperature inserts with superconducting magnets for the low field range (0-9 T) and in resistive magnets in the Chinese High Magnetic Field Laboratory (CHMFL, China) and the National High Magnetic Field Laboratory (NHMFL, USA). The temperature dependence of the resistivity down to 50 mK was measured using a dilution refrigerator. The CaAs$_3$ crystals were contacted to the chip carrier using silver paints and gold wires. The typical channel size of the measured devices in transport experiments is 1.5 (length)×1.5 (width)×0.3 (thickness) mm$^3$ (Sample C1). The resistivity was measured using either the *ac* lock-in technique or the *dc* method with a current in the range of 1~100 nA applied to the sample while monitoring the voltage in the four-terminal configuration.

The Seebeck coefficient was measured in physical property measurement system (Quantum Design) by a home-built device. The CaAs$_3$ crystals were contacted to the chip carrier using silver paints and gold wires, Type-E thermocouple (Omega) was used to measure temperature gradient and Nanovoltmeter 2182A (Keithley) was used to measure thermoelectricity voltage.

**Magnetic torque measurement.** The magnetic torque measurements were performed using capacitive cantilever in resistive magnets in CHMFL. In order to estimate the background signal from the cantilever and the cable, an empty cantilever was calibrated on the same experimental condition. To be noted, 0° (or 90°) is not high-symmetry crystalline axes of triclinic CaAs$_3$.

**Magneto-infrared measurement.** Magneto-infrared spectrum was measured in the mid-infrared range in the Faraday geometry. The spectra up to 17.5 T were measured with a superconducting magnet at NHMFL. The spectra up to 30 T were measured with a resistive magnet at Laboratoire National des Champs Magnétiques Intenses (LNCMI), Grenoble. The IR beam from the FT-IR spectrometer propagates in a vacuum and is shed on the sample at a near-incident angle. The reflected beam was detected by a 4.2 K composite silicon bolometer.

**Density functional theory (DFT) calculations.** DFT calculations were carried out using the Quantum Espresso package.[71] The Perdew-Burke-Ernzerhof form of the exchange-correlation functional was used[72] and spin-orbit coupling was included. Experimentally reported structure for CaAs$_3$ was used. A plane wave cutoff of 40 Ry was employed, and the Brillouin zone was sampled using an 8×8×8 Monkhorst-Pack *k*-point grid[73]. A dense grid of 16×16×16 *k*-points was chosen for density of states computations.

**Acknowledgments**

We thank Hai-Zhou Lu, Hua Jiang, Haiwen Liu, Yijia Wu, and Zhongbo Yan for stimulating discussions, Ran Tao and Tong Zhang for assistance with crystal characterization. C.Z. was sponsored by the National Key Research and Development Program of China (Grant No. 2022YFA1405700), the National Natural Science Foundation of China (Grant No. 12174069, 92365104), and Shuguang Program from the Shanghai Education Development Foundation. F.X. was supported by the National Natural Science Foundation of China (52225207, 52150103, 11934005, and 11874116), the Science and Technology Commission of Shanghai (Grant No. 19511120500), the Shanghai Municipal Science and Technology Major Project (Grant No. 2019SHZDZX01), the Program of Shanghai Academic/Technology Research Leader (Grant No. 20XD1400200), and Shanghai Pilot Program for Basic Research - Fudan University 21TQ1400100 (21TQ006). X.Y. was sponsored by the National Key R&D Program of China (Grant No. 2023YFA1407500) and the National Natural Science Foundation of China (Grants No. 12174104 and No. 62005079). Part of the sample fabrication was performed at Fudan Nano-fabrication Laboratory. Part of transport measurements was performed at the High Magnetic Field Laboratory,

CAS. A portion of this work was performed at the National High Magnetic Field Laboratory (USA), which is supported by National Science Foundation Cooperative Agreement No. DMR-1644779 and the State of Florida. J.Z. was supported by the National Key R&D Program of the MOST of China (Grant No. 2022YFA1602603), and the Natural Science Foundation of China (Grants No. 12122411). The work at Nanjing University was supported by the National Key Research and Development Program of China with Grant No. 2021YFA1400400, and the National Natural Science Foundation of China with Grant No. 12074174. N.B.J. is supported by the Prime Ministers Research Fellowship. A.N. acknowledges support from the start-up grant at the Indian Institute of Science (SG/MHRD-19-0001) and DST-SERB (project number SRG/2020/000153).

**Author contributions**

C.Z. and F.X. conceived the ideas and supervised the overall research. S.L. and J.W. synthetized $CaAs_3$ crystals. C.Z., Y.W., J.Z. M.Z., Y.M., Y.Z., Y.Y., P.L., Y.Z., and X.K. carried out the electric transport measurements. L.X. and Z.Z. performed the thermoelectric transport measurement. W.W. and X.Y. carried out the magneto-infrared experiments. J.Z., A.S., M.O., J.W., M.O., and L.P. helped with the transport, torque, and infrared experiments at high magnetic fields. W.J., F.Z., and D.Q. conducted the photoemission experiment. C.Z. and Y.W. analyzed the data. N.B.J., Y.Z., and A.N. provided theoretical support. C.Z., Y.W., X.Y., and F.X. wrote the paper with help from all other co-authors.

**Competing financial interests**

The authors declare no competing financial interests.

**Data availability**

All data that support the findings of this study are available from the corresponding authors upon reasonable request.

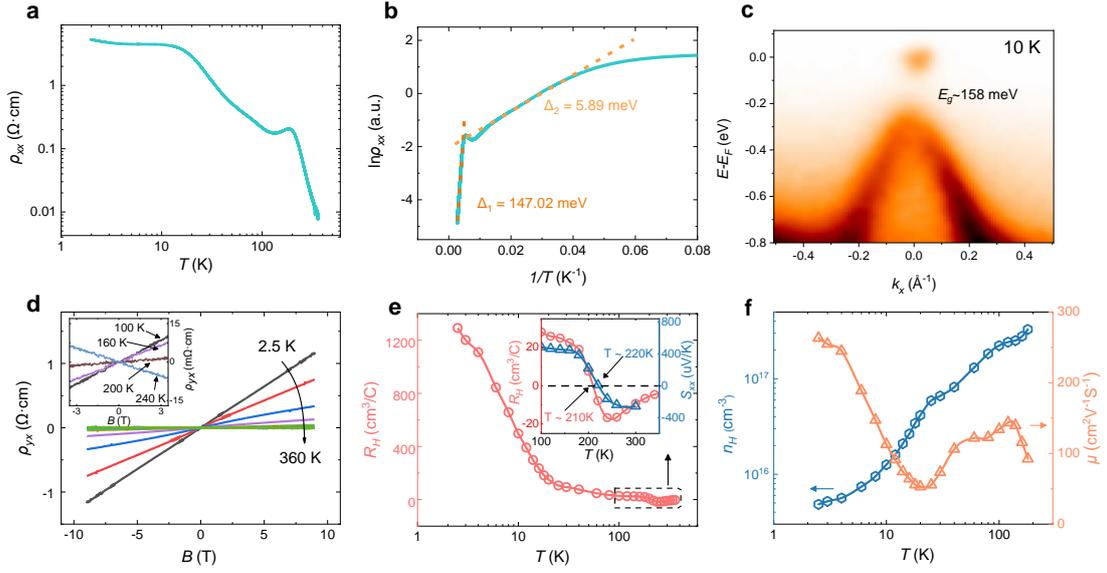

**Fig. 1 | Transport results and ARPES data of CaAs₃ crystals. a**, Longitudinal resistivity $\rho_{xx}$ of CaAs₃ plotted as a function of temperature (measured in the *a-c* plane). **b**, Arrhenius plot of $\rho_{xx}$ with two activation regimes (dashed lines) and the corresponding activation energies. **c**, Band dispersion at 10 K measured by ARPES with the photon energy of 98 eV. **d**, Hall resistivity curves from 360 K to 2.5 K with the magnetic field perpendicular to the *a-c* plane. The inset is an enlarged view of Hall resistivity curves in the vicinity of the sign reversal. **e**, Hall coefficient $R_H$ (derived from the linear fitting of Hall resistivity from -9 T to 9 T) plotted as a function of temperatures. The inset is the comparison of the Hall coefficient $R_H$ (red circle) and the Seebeck coefficient $S_{xx}$ (blue triangle), showing similar sign-reversal temperature. **f**, The temperature dependence of carrier density (blue hexagon) and mobility (yellow triangle) below the sign reversal temperature of $R_H$.

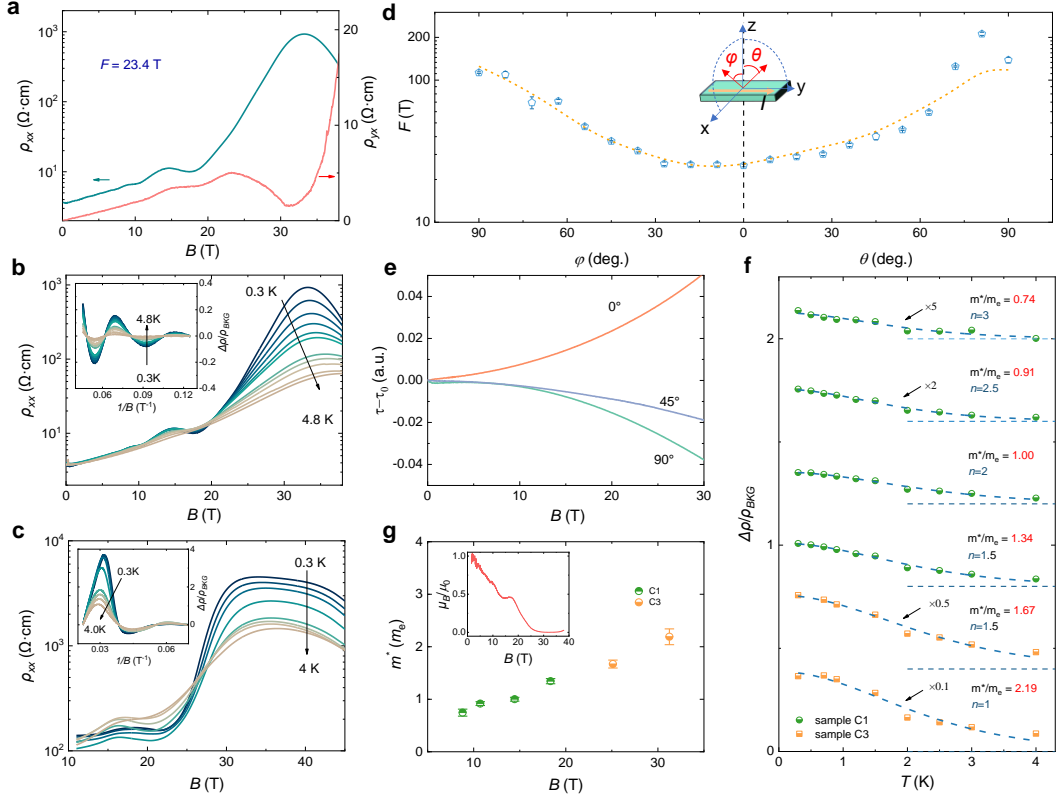

**Fig. 2 | Quantum oscillations observed in high magnetic fields. a**, The magnetic field dependence of $\rho_{xx}$ and $\rho_{yx}$ at 0.3 K with prominent SdH oscillations (sample C1). The magnetic field is applied perpendicular to the *a-c* plane. **b-c**, MR curves of sample C1 and C3 at different temperatures from 0.3 K to 4.8 K. The inset is the extracted oscillatory component normalized by $\rho_{BKG}$ at different temperatures. **d**, The angle dependence of SdH oscillation frequencies of sample C3. Pentagon dot and dashed line represent measured data and fitting curve based on the Fermi surface anisotropy given by DFT calculations, respectively. **e**, Torque measurement of sample C3 at different angles at 1.4 K. 0° (90°) represents the magnetic field perpendicular (parallel) to *a-c* plane. **f**, The fitting of the relative resistance change by the temperature factor $R_T$ of the LK formula of sample C1 and C3 at different fields. Note that it is a stacked view with blue dashed lines representing the offset values. The effective mass fitting parameter is shown near each curve. **g**, The extracted effective mass values of sample C1 and C3 at different magnetic fields. The inset is the ratio of the mobility values ($\mu_B/\mu_0$) in magnetic fields calculated from the Hall angle, where $\mu_B$ and $\mu_0$ are the mobility values at a magnetic field of *B* and zero field, respectively.

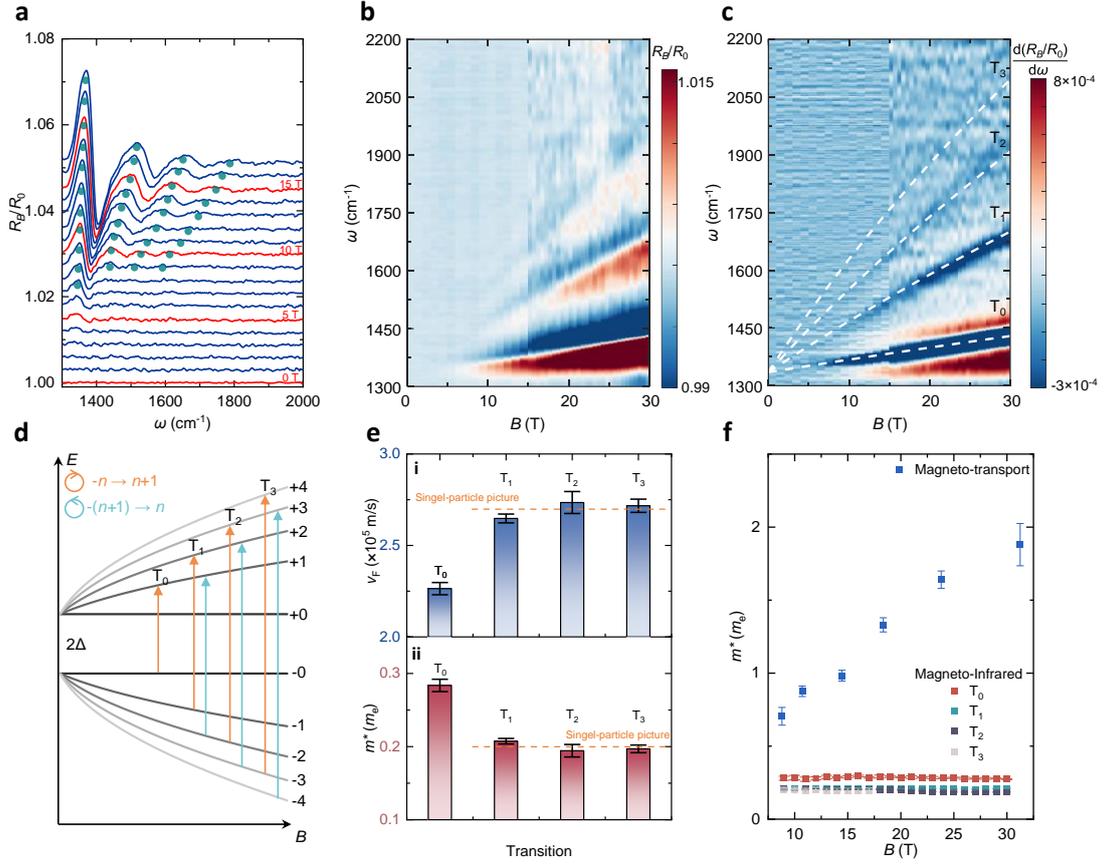

**Fig. 3 | Magneto-infrared spectroscopy of CaAs$_3$. a,** Relative magneto-reflectivity spectra $R_B/R_0$ measured in a superconducting magnet up to 17 T. All curves are vertically stacked for clarification with a series of interband-Landau-level transitions. The apparent peak features are denoted by deep green dots. **b,** False-color plot of relative magneto-reflectivity spectra in the whole field regime. The spectra for the higher magnetic field are measured in a resistivity magnet. **c,** False-color plot of the first-order derivative of relative magneto-reflectivity spectra d($R_B/R_0$)/d$\omega$, where the minima are associated with the optical excitation energy. The white dashed curves are the fitting results based on the massive Dirac model. **d**, The schematic Landau level spectrum of CaAs$_3$. The orange and cyan arrows denote the interband-Landau-level transitions with opposite circular polarization. **e**. Fermi velocity (panel i) and effective mass (panel ii) fitted from the separate interband-Landau-level transitions. Dashed orange lines show the Fermi velocity and effective mass obtained from the fitting to a single-particle model. **f**. The comparison between effective mass extracted from the quantum oscillations (blue square) and interband-Landau-level transition T$_0$ (red square). The values of effective mass obtained from the optical transitions T$_1$ (deep green square), T$_2$ (deep gray square) and T$_3$ (light gray square) are less than that from T$_0$.
17 / 18

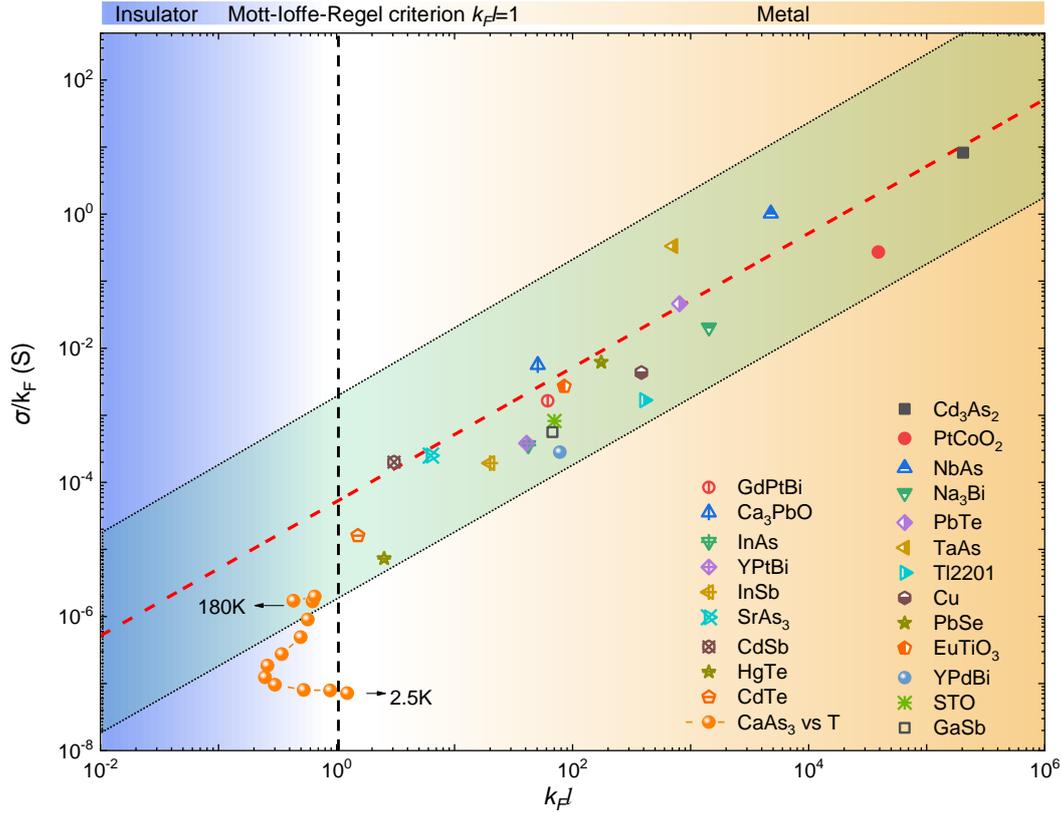

**Fig. 4 | The scaling plot of $\sigma/k_F$ against the metallicity parameter $k_F l$.** The red dashed line is the theoretical curve given by the Drude model. The yellow-white-blue gradient background corresponds to the region with $k_F l \gg 1$, $k_F l \sim 1$, $k_F l \ll 1$. The black dashed line marks the Mott-Ioffe-Regel limit $k_F l = 1$. The Fermi wave vector $k_F$ was derived from quantum oscillation by the Onsager relation and the mean free path $l$ was derived from the Fermi wave vector $k_F$ and the Hall mobility $\mu$ through $l = \hbar k_F \mu / e$.

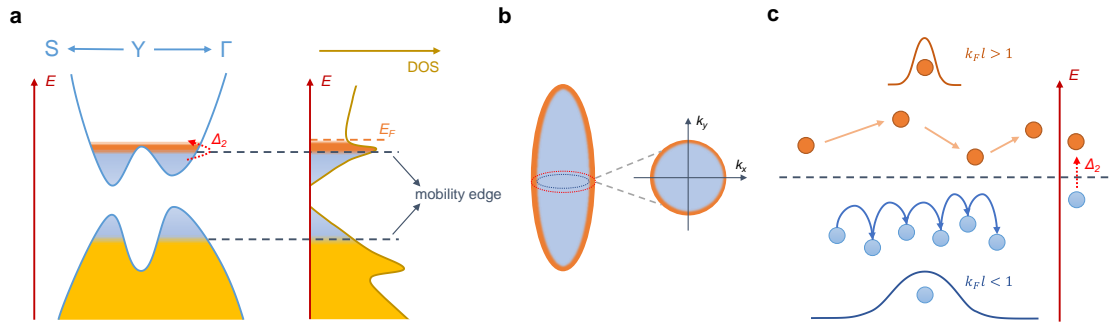

**Fig. 5 | Schematic plots of the electronic structure and carrier transport dynamics of CaAs$_3$.** **a**, Sketch of the band dispersion (left) and DOS (right) in CaAs$_3$. The yellow and light blue regions separated by the mobility edge represent the extended and localized states, respectively, and the orange region corresponds to conducting carriers of the extended states near the Fermi energy $E_F$. **b**, Sketch of the Fermi sphere. **c**, The transport dynamics of the extended ($k_F l > 1$) and localized ($k_F l < 1$) states, respectively. The dashed line corresponds to the mobility edge. The red dashed arrows in **a** and **c** represent the excitation from the localized state to the extended state characterized by $\Delta_2$.